\documentclass{tlp}
\usepackage{aopmath}

\usepackage{xspace}
\usepackage{graphicx}
\usepackage{verbatim}
\usepackage{url}
\newcommand{\toolName}[1]{#1\xspace}
\newcommand{\DAHL}{\toolName{DAHL}}
\newcommand{\Ptwo}{\toolName{P2}}
\newcommand{\Mace}{\toolName{Mace}}

\newcommand{\libevent}{\toolName{libevent}}
\newcommand{\SicstusProlog}{\toolName{SICStus Prolog}}
\newcommand{\Prolog}{\toolName{Prolog}}

\newcommand{\codefontstyle}{\normalsize}
\newcommand{\codeindent}{2ex}

\newcommand{\meta}[1]{{\normalfont\slshape#1}}

\newcommand{\ttcode}{\ttfamily\def\_{\string_}}

\newenvironment{code*}[1][\codefontstyle]{\block#1\verbatim}{\endverbatim\endblock}
\newenvironment{code}[1][\codefontstyle]{\block\ttcode #1}{\endblock}

\makeatletter
\newenvironment{block}
               {\list{}{\leftmargin \codeindent
                        \topsep 2pt}%
                \item\relax}
               {\endlist}

\makeatother

\newcommand{\vertcommand}[1]{{\ttcode #1}}
\makeatletter
\def\@makevertactive{\catcode`|=\active}
{
\@makevertactive
\gdef\activevert{\@makevertactive\def|##1|{\vertcommand{##1}}}%
}
\makeatother

\newcommand{\constF}{F}

\renewcommand{\codeindent}{2em}

\title{Applying Prolog to Develop Distributed Systems}

\author[N. Lopes, J. Navarro, A. Rybalchenko and A. Singh]{
  Nuno P. Lopes$^\ast$, Juan A. Navarro$^\dagger$, Andrey Rybalchenko$^\dagger$,
  and Atul Singh$^\ddagger$ \\
  $^\ast$ INESC-ID /
  Instituto Superior T\'ecnico, Technical University of Lisbon\\ \\
  $^\dagger$ Technische Universit\"at M\"unchen \\ \\
  $^\ddagger$ NEC Research Labs, 
  Princeton, NJ}

\begin{document}

\maketitle

\vspace{1ex}
\noindent
Note: This article has been published in
\emph{Theory and Practice of Logic Programming},
26th Int'l. Conference on Logic Programming (ICLP'10) Special Issue,
10(4-6):691-707, July 2010,
\copyright Cambridge University Press.
\vspace{4ex}

\begin{abstract} 
  Development of distributed systems is a difficult task.
Declarative programming techniques hold a promising potential for
effectively supporting programmer in this challenge.
While Datalog-based languages have been actively explored for
programming distributed systems, Prolog received relatively little
attention in this application area so far.
In this paper we present a Prolog-based programming system, called
DAHL, for the declarative development of distributed systems.
DAHL extends Prolog with an event-driven control mechanism
and built-in networking procedures.
Our experimental evaluation using a distributed hash-table data
structure, a protocol for achieving Byzantine fault tolerance, and a
distributed software model checker -- all implemented in DAHL -- indicates
the viability of the approach.


\end{abstract}

\section{Introduction}

Declarative Networking is a promising direction in the quest for
distributed programming systems that meet the challenges of building
reliable and efficient distributed applications~\cite{LooSIGOPS05}. 
As the name suggests, Declarative Networking advocates a high-level
programming paradigm where the programmer specifies what has to be
computed and communicated over the network, and then the compiler
translates the specification into executable code.
Its main applications are various network protocols, including sensor
networks~\cite{Sensors}, fault tolerance
protocols~\cite{SinghNSDI08,AlvaroNetDB09},
distributed hash tables~\cite{LooSIGOPS05}, and data
replication systems~\cite{PadreTech08}.

Current implementations of Declarative Networking adapt Datalog for
the domain of networking applications. 
The resulting programming languages have a bottom-up evaluation
semantics where 
the evaluation of (Datalog) clauses causes the execution
of corresponding networking actions.
Since Datalog is a not a general purpose programming language, its
adaptation for Declarative Networking required 
a reformulation of the language to allow the developer some control
over the flow of execution, to make available
expressive data types, and to maintain a mutable state.
Programmers often include C/C++ fragments on ordinary
occasions~\cite{SinghNSDI08}, while research efforts are invested to
better couple the required additional features with the Datalog
evaluation model~\cite{Mao09,BOOM}.

In this paper we explore the applicability of Prolog as a basis for
Declarative Networking.
In contrast to Datalog, Prolog is a general purpose programming
language.
Since Prolog is considered to be a practical tool for programming in
logic, its adaptation to distributed programming can focus only on the
networked communication aspects.
In the process, we put Prolog into an event-driven execution
environment, where each node interprets messages received from the
network as queries on its own local database, and provide a collection
of procedures for communication via message passing.
As a result, we obtain an extension of Prolog that can be applied for
distributed programming.
Its implementation, called \DAHL\footnote{Available at:
\url{http://www7.in.tum.de/tools/dahl/}}, consists of a bytecode
compiler and a runtime environment.
\DAHL builds upon an existing Prolog infrastructure~\cite{sicstus} and a
networking library~\cite{libevent}.

We evaluate \DAHL on a range of distributed applications including the
Chord distributed hash table~\cite{Chord}, the Byzantine fault
tolerance protocol Zyzzyva~\cite{KotlaSIGOPS07}, and a distributed
software model checker~\cite{darmc}.
\DAHL implementations are comparable to existing Declarative Networking
approaches in terms of succinctness,
and do not require any C/C++ workarounds.
Moreover, we also show that \DAHL's performance is competitive
with C++ runtimes produced with Mace, a tool that supports the
development of robust distributed systems \cite{KillianPLDI2007}, 
while significantly reducing code size.

In this paper we present the following contributions:
\begin{itemize}
\item We demonstrate that Prolog is a suitable basis for the
design of a programming language for Declarative Networking.
Our approach exploits Prolog's strengths to provide
general purpose programming features, while retaining
its conceptual ties with the declarative paradigm.
\item We provide an efficient and robust programming system for DAHL
that includes a compiler and a runtime environment.
\item We demonstrate the practicality of DAHL via an experimental
evaluation on a range of distributed applications. 
\end{itemize}

We organize the paper in the following way:
In Section~\ref{sec:examples}, we introduce \DAHL using a simple
spanning-tree protocol as example.
The programmer interface that allows the
development of distributed applications is described in
Section~\ref{sec:writing}.
We present implementation details of \DAHL in
Section~\ref{sec:implementation}, and evaluation results in
Section~\ref{sec:evaluation}. We also give
a review of the related work in Section~\ref{sec:related},
and then conclude in Section~\ref{sec:conclusion}.


\section{\DAHL by example}
\label{sec:examples}


\begin{figure}
\begin{code*}[\small]
:- event span_tree/2.

span_tree(Root, Parent) :-
  \+ tree(Root, _),
  assert(tree(Root, Parent)),
  this_node(ThisAddr),
  sendall(
    Node,
    neighbor(Node),
    span_tree(Root, ThisAddr)
  ).
\end{code*}
  \caption{\DAHL program to compute a spanning-tree overlay.}
  \label{fig-span-tree-dahl}
\end{figure}

\newcommand{\msgSpanTree}[2]{\texttt{span\_tree(#1,#2)}\xspace}
\newcommand{\factTree}[2]{\texttt{tree(#1,#2)}\xspace}
\newcommand{\factNeighbor}[1]{\texttt{neighbor(#1)}\xspace}
\newcommand{\nodeRoot}{{\normalfont\slshape Root}}
\newcommand{\nodeN}{{\normalfont\slshape Node}}
\newcommand{\thisAddr}{{\normalfont\slshape ThisAddr}}

In this section, we illustrate \DAHL by using an example program that
implements a simple protocol for constructing a spanning-tree overlay
in a computer network.
Tree-based overlay networks have received significant attention from the
academic community~\cite{CastroSOSP03,ChuUSENIX04,JannottiOSDI00,BanerjeeSIGCOMM02} and have also
seen successful commercial deployment~\cite{LiJSAC2007}.
In these tree overlays, after some network node has been selected to be the root
node, we require that each other node is able to forward messages to the
root node. 
After the spanning-tree overlay is constructed, each node can send a
message to the root by either using a direct link, if available, or
relying on the ability of some neighbor to forward messages to the
root.
If a node is not connected to the root via a sequence of links then
the node cannot send any messages to the root.

The overlay is constructed by propagating among the network nodes the
information on how to forward to the root node. 
This information is given by the address of the next node towards the
root.
We assume that initially each node stores the addresses of its
immediate neighbors in the (local) database.
This information is loaded at startup by each node (e.g., at the command
line or from a configuration file) into the \factNeighbor{\nodeN} table.

A node can directly access its neighbors by sending messages over the
corresponding network links.
At the initial step of the overlay construction, the designated root
node, say \nodeRoot, is triggered by sending it a
message~\msgSpanTree{\nodeRoot}{\nodeRoot}.
Then, the root node sends \msgSpanTree{\nodeRoot}{\nodeRoot} to
each neighbor node. 
At a neighbor, say \nodeN, this message leads to the addition of the
fact \factTree{\nodeRoot}{\nodeRoot} to the database, thus,
recording the possibility of reaching the tree root in a single step.
Furthermore, \nodeN propagates this information to its neighbors by
sending a message~\msgSpanTree{\nodeRoot}{\nodeN}.
Upon reception, each \nodeN's neighbor adds
\factTree{\nodeRoot}{\nodeN} to its database and continues the
propagation.

Our implementation of the spanning-tree protocol relies on a
combination of Prolog with networking and distribution-specific
extensions to achieve the goal, see Figure~\ref{fig-span-tree-dahl}.
When the initial message \msgSpanTree{\nodeRoot}{\nodeRoot}
arrives at \nodeRoot, it is interpreted as a Prolog query. 
The query execution is carried out by the corresponding procedure
\texttt{span\_tree/2}, which is authorized to execute queries that arrive
from the network due to the declaration~\texttt{event span\_tree/2}.
The procedure \texttt{span\_tree/2} uses standard Prolog predicates as
well as our extensions.
First, \texttt{span\_tree/2} checks if the information how to reach
the root node is already available.
If it is the case, the execution of the procedure fails, and since the
initiating query was issued by the network, \DAHL ignores the failure
and continues with the next message as soon as it arrives.
Otherwise, a fact recording the root's reachability is added to the
database.
We propagate the corresponding information to the neighbors, whose
addresses are stored by each node as facts~\texttt{neighbor/1} in the
database.
The message that is sent to each neighbor contains the sender address,
which is required for the overlay construction.
We obtain this address by using a \DAHL built-in
predicate~\texttt{this\_node/1}.
The communication with the neighbors is implemented using a \DAHL
built-in procedure \texttt{sendall/3}, which is inspired by the ``all
solutions'' predicates provided by Prolog, e.g.,~\texttt{findall/3} or
\texttt{setof/3}.
For each address that can be bound to \nodeN\ by evaluating
\factNeighbor{\nodeN}, the execution of \texttt{sendall} sends a
message \msgSpanTree{\nodeRoot}{\thisAddr}, i.e., the message is sent
to all neighbors.

In summary, our example shows that we can apply Prolog for developing
distributed protocols by putting it into an event-driven execution
environment and by extending the standard library with
networking-specific built-in procedures. 
A more complex example is shown in Figure~\ref{fig:zyzzyva}, which is an excerpt
of our implementation of the Zyzzyva Byzantine fault tolerant protocol.
In the rest of the paper, we briefly introduce the extensions and
describe their interplay with \Prolog for implementing a distributed
hash-table data structure, a protocol achieving the Byzantine fault
tolerance, and a distributed version of a software model-checking
algorithm.


\section{Programming interface for distributed applications}
\label{sec:writing}

We now present the interface
for developers to implement distributed applications.
The interface consists of an event driven control and
a set of primitives to send messages over the network.
Our implementation of this interface is described later in
Section~\ref{sec:implementation}.

\begingroup\activevert
\paragraph{Messages and event handlers}

Nodes communicate by exchanging messages represented
by Prolog terms.
When a message is received by a node, it triggers
the evaluation of the matching event handler.
An event handler corresponds to a Prolog predicate definition
and its evaluation is done as a Prolog query.

The declaration
\begin{code}
:- event \meta{PredSpec}, ..., \meta{PredSpec}.
\end{code}
turns each predicate specified by \meta{PredSpec}
into an event handler for messages that match its specification.
A predicate specification is an expression of the form
|p/n| where |p| is a predicate name and |n| its arity.
For example,
\begin{code}
:- event q/2. \\[1em]
q(X, Y) :- \meta{Body}.
\end{code}
declares the |q/2| predicate as the event handler for
messages of the form |q(X, Y)|.
In other words, the |event| declaration allows the evaluation
of a predicate to be triggered when a matching message is
received from the network.

In a running application, a node waits until a message is
received from the network.
When a message is received, and if the corresponding predicate is
declared as an |event|, Prolog's standard evaluation strategy is used
to compute the first solution to the message as if it was posed as a
query.
As the query is evaluated, the event handler can modify the local
state of the node, e.g., with |assert|/|retract|, or produce messages
to send to other nodes.
The solution to this query, or the failure to find a solution, is
discarded, but the side effects of the evaluation are not.
Messages that are not declared as events are also discarded.
Event handlers triggered by different messages are evaluated
atomically in sequence, i.e., the evaluation of a new message does not
start until the evaluation of the previous one has finished.
Atomic evaluation avoids concurrency issues that could arise when
processing multiple messages at once.

\bigskip
\DAHL provides the |send/2| and |sendall/3| built-in predicates to send
messages over the network. The predicate
\begin{code}
send(\meta{Address}, \meta{Message})
\end{code}
sends \meta{Message} to the node at \meta{Address}.
Evaluation of the predicate succeeds as soon as the underlying transport
protocol reports the message as sent, and evaluation of the rest of
the query continues. If an error occurs (e.g., \emph{Address}
is unreachable), the predicate fails and backtracks, e.g., to find an
alternate destination.
This is the default behavior and can be configured to throw exceptions
or ignore errors instead.

Low level details, such as opening and closing network connections,
are abstracted away and handled
automatically by the \DAHL runtime.
If needed, developers can also access low level
primitives to open/close connections themselves.

\bigskip
Multiple messages can be sent using
\begin{code}
sendall(\meta{Address}, \meta{Generator}, \meta{Message})
\end{code}
which, for every solution of \meta{Generator}, sends a
\meta{Message} to \meta{Address}.
A developer can use this predicate to broadcast
a message to all neighbors of a node. For example,
\begin{code}
sendall(N, neighbor(N), ping)
\end{code}
sends a |ping| message to every node |N| which is a solution to
|neighbor(N)|.
Moreover, both the \meta{Address} and the \meta{Message}
of the |sendall| operation can be determined by the \meta{Generator}.
For example,
\begin{code}
sendall(N, (task(T), assign(T, N)), solve(T)).
\end{code}
distributes a number of tasks among a set of nodes.

Low level implementations can optimize for particular usages of
|sendall|. As an example, when \emph{Message} does not
depend on \emph{Generator}, a network-level multicast/broadcast protocol
can be used to provide a more efficient operation.

\bigskip
Another feature provided by the \DAHL interface is that of alarms.
Alarms are used by nodes to cause the evaluation of a local
event handler at a specified time in the future.
Similar to events, the declaration
\begin{code}
:- alarm \meta{PredSpec}, ..., \meta{PredSpec}.
\end{code}
turns each predicate specified by \meta{PredSpec} into an alarm handler.
The predicate
\begin{code}
alarm(\emph{Message}, \emph{MSecs})
\end{code}
succeeds after setting up a reminder to insert \emph{Message}
in the local queue after \meta{MSecs} have elapsed.
The |alarm/2| predicate can also be used to trigger event
handlers declared with the |event| directive; but a predicate declared
as |alarm| will never respond to messages
from the network (e.g., produced with |send| or |sendall|).

\paragraph{Authentication}

When running a distributed protocol over an untrusted network,
it is often required for messages to be signed in order to
authenticate their origin.
\DAHL's interface allows an application to be easily augmented with
authentication by replacing |send/2| with the predicate
\begin{code}
send\_signed(\meta{Address}, \meta{Message})
\end{code}
that attaches authentication metadata to the \meta{Message}
sent to \emph{Address}.
Similarly, the |sendall\_signed/3| predicate, analogous to |sendall/3|,
is provided.

On the receiving end, the predicates
\begin{code}
signed\_by(\meta{Address}, \meta{Signature}) \\
signed\_by(\meta{Address}) \\
signed
\end{code}
check on demand whether the incoming message
(and whose event handler is being evaluated)
was properly signed.
Additionally, if present, \meta{Address} is unified
with the address of the sender and \meta{Signature} with the signature
metadata.
If the message was not signed, or had an invalid
signature, these predicates fail.

Since cryptographic operations are often computing intensive,
these predicates allow the programmer to schedule the
validation of signatures at an appropriate time
in the evaluation of an event handler. For example,
\begin{code}
request(Req) :- valid(Req), signed\_by(Addr), ...
\end{code}
checks the validity of a request before performing a, possibly more
expensive, validation of the signature. This strategy is
applied in the definition of |request/1| in
our implementation of Zyzzyva (Figure~\ref{fig:zyzzyva}).

Authenticity in \DAHL is based
on OpenSSL's implementation of HMAC for signing messages
and MD5 for computing message digests. Alternative 
crypto-algorithms can be selected and accessed
through the same high-level interface. 

\endgroup


\section{Implementation}
\label{sec:implementation}



\begin{figure}
\includegraphics[scale=0.5]{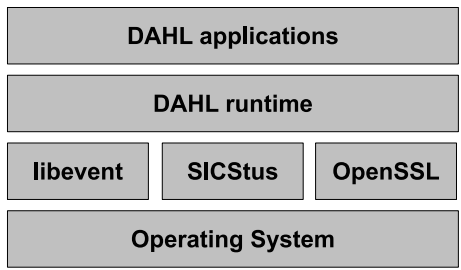}
\caption{The \DAHL software stack.}
\label{fig:dahl_stack}
\end{figure}

The software architecture of \DAHL is shown in Figure~\ref{fig:dahl_stack}.
It consists of a \DAHL compiler (based on \SicstusProlog compiler from
\citeNP{sicstus}), a high-performance event dispatching library
(\libevent from \citeNP{libevent}), the OpenSSL library to provide the
cryptographic primitives in the language, the \DAHL runtime, and \DAHL applications.
We use the off-the-shelf \SicstusProlog compiler to quickly build the \DAHL
system and utilize its industrial-strength performance and robustness for
achieving high performance.
We do not describe the details of how we interfaced \libevent and OpenSSL
since they are standard, instead we describe in detail the novel
aspects of \DAHL: how the runtime works, some optimizations that were
implemented, and the networking aspects.

\paragraph{Runtime}
\DAHL's runtime consists of a library written in \Prolog (with around 460 lines
of code), which implements the built-in predicates, and a networking back-end
written in C (around 450 lines).
It is the networking back-end that interfaces with both \libevent and OpenSSL.
This back-end interfaces with \Prolog through stubs generated automatically by
the \SicstusProlog compiler.

\DAHL programs are interpreted directly by the \SicstusProlog compiler, but
under the \DAHL runtime control.
The main program in execution is a loop that is part of the runtime, and a \DAHL
program's code is only called when an appropriate event arrives from the
network, or when a timer is triggered.
Those events are processed by libevent.

\begin{figure}
\includegraphics[scale=0.4]{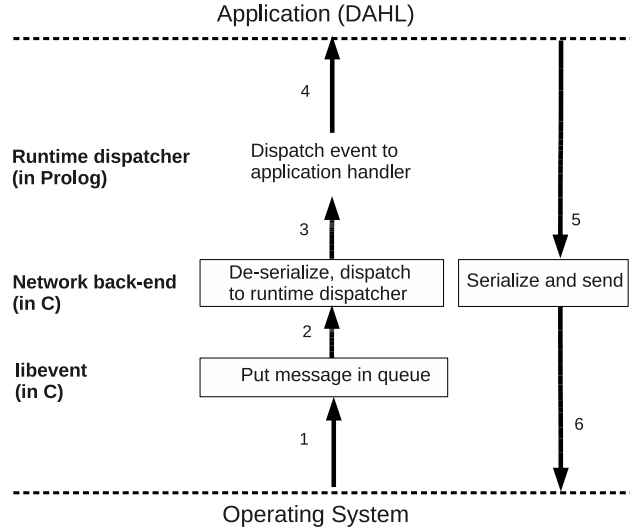}
\caption{Internals of \DAHL runtime shown by tracing the flow of message processing.}
\label{fig:dahl_msg_dispatch}
\end{figure}

Figure~\ref{fig:dahl_msg_dispatch} shows the execution flow for processing a
message that arrives from the network (steps 1--4), and for a message that
is sent from an application (steps 5--6) in more detail.
When a message arrives from the network, the operating system dispatches
it to \libevent (step 1), which queues the message.
Then, when the \DAHL runtime asks for the next message, \libevent picks one
arbitrarily and delivers it to the \DAHL network
back-end (step 2). The \DAHL network back-end then deserializes the message and
calls the runtime dispatcher (in \Prolog) through a stub (step 3). Finally,
the dispatcher calls the corresponding event handler of the
application (step 4).
When a \DAHL application sends a message, the message is first handed over to the
\DAHL runtime through a stub (step 5). The runtime then serializes the message
and delegates the network transmission to the operating system (step 6).

\paragraph{Optimizations} We implemented several optimizations in the \DAHL runtime
to improve its performance. Here, we present these optimizations
in detail. The deserialization of network messages was a CPU-intensive operation 
since the \SicstusProlog compiler implements this operation in \Prolog through
a complex process chain.
Since each message sent was serialized to a single atom, it led to an
explosion in memory usage because the \SicstusProlog compiler aggressively
caches all atoms.
We therefore implemented our own custom deserialization in C to improve the
performance.
This resulted in a performance improvement of the deserialization function of
about 70\%.

As described before, the main loop is implemented in \Prolog, and it calls
a function in C that ``produces'' events through \libevent, which are then
dispatched from within the \Prolog environment.
The loop is implemented as a \Prolog rule that first calls the
external C function, and then fails and backtracks to the beginning
of the rule in order to iterate.
This provides an important advantage, which is that every event/alarm handler
is executed in a ``clean'' environment, as all the garbage possibly left
by a previous handler is discarded.
Moreover, it improves the performance of the garbage collector (GC), as the \SicstusProlog
compiler will delete most of such garbage when backtracking as an optimization,
reducing the overhead of the GC.
Our tests show that without this environment cleanup, the overhead of the GC
would be noticeable (from 8\% to 45\%).

\paragraph{Network Support} Currently all the network messages are sent using the TCP protocol,
which requires establishing a connection before the first contact.
The \DAHL runtime automatically establishes these connections when needed, and caches
them indefinitely for future contacts. It is straightforward to replace TCP with UDP,
though the application needs to have mechanisms to handle message loss.


\section{Evaluation}
\label{sec:evaluation} 

In this section, we present an evaluation of \DAHL in terms of run-time
performance, language expressiveness, and succinctness of programs.
Implementations of networking protocols, like
Chord~\cite{Chord} and Zyzzyva~\cite{KotlaSIGOPS07}, as well as CPU-bound
applications like D'ARMC~\cite{darmc} demonstrate the applicability
of \DAHL in the development of real-life and complex systems.
We compare the results with alternative implementations of these
protocols in P2~\cite{LooSIGMOD06}, the original implementation
of Declarative Networking, and \Mace~\cite{KillianPLDI2007},
an extension of C++ with networking capabilities
and a state-machine specification language.

\subsection{Raw Performance}

To evaluate the performance of the \DAHL runtime, we performed a test to compare
the performance of \Ptwo, \Mace, C, and \DAHL.
We performed a simple network ping-pong experiment.
One of the machines (called a client) sends a small 20-byte `ping' message to
the other machine (called a server) which immediately responds with a small
20-byte `pong' message.
We used as many client machines as needed to saturate the server in order to
measure its raw throughput.
The measurement of the number of requests served per second was done at the
server.
The machines were connected by a gigabit switch with a round trip latency of
0.09 ms, and both the network and the machines were unloaded.
The results are presented in Table~\ref{table:compare_raw_ping_pong}.

\begin{table}
\caption{Comparing raw network performance of \Ptwo, \DAHL, \Mace, \Mace
compiled with `-O2' optimizations, and plain C as the maximum number of pings
responded by the server in a second.}
\begin{tabular}{ccccc}
\hline\hline
\Ptwo & \DAHL & \Mace   & \Mace (w/ -O2) & C\\
\hline
230 &  14,000 & 14,221  & 21,937 & 142,800\\
\hline\hline
\end{tabular}
\label{table:compare_raw_ping_pong}
\end{table}

\def\NoteAnalysis{%
by providing higher level programming abstractions
which are also more amenable to static analysis and
program verification techniques}

First, we note that the \DAHL runtime outperforms \Ptwo's performance.
We believe that the reason behind \Ptwo's poor performance is that the runtime
of \Ptwo is not yet optimized while \DAHL uses the SICStus 4 compiler
that has been already optimized.
Second, \DAHL  is as fast as \Mace.
However, given that \Mace is a restricted form of C++, it can exploit powerful
C++ optimizing compilers.
For example, with the `-O2' set of optimizations of gcc 4.1, \Mace's performance
improves by 60\%.
As an upper bound on the performance, we also present the performance of a C
implementation and note that all the systems that strive to improve the analysis capability---\NoteAnalysis---are an order of magnitude slower.

\subsection{Chord}
In this subsection, we evaluate the performance of an implementation of Chord
(a distributed hash table) in \DAHL.
Our implementation of Chord implements all features detailed in the original
paper~\cite{Chord}.
To compare with the P2 Chord implementation, we obtained the latest release of
P2.\footnote{Version 3570 in \url{https://svn.declarativity.net/p2/trunk}.}
Unfortunately, we were unable to get P2 Chord running in our local setup.
We therefore cite results from their paper~\cite{LooSIGOPS05}. 

\paragraph{Setup}
We used ModelNet~\cite{modelnet} to emulate a GT-ITM transit-stub topology
consisting of 100 to 500 stubs and ten transit nodes.
The stub-transit links had a latency of 2 ms and 10 Mbps of bandwidth, while
transit-transit links had a latency of 50 ms and 100 Mbps of bandwidth.
We used 10 physical hosts, each with dual-core AMD Opteron 2.6 GHz processor
with 8GB of RAM, and running Linux kernel version 2.6.24.
We ran 10 to 50 virtual nodes on each physical node, producing a population of
100 to 500 nodes.
In each experiment, neither the CPU nor the RAM were the bottleneck.
This setup reproduces the topology used by the \Ptwo experiments
in~\cite{LooSIGOPS05}, although they used Emulab~\cite{emulab}.

\paragraph{Static Membership} Our first goal was to see if the \DAHL
implementation met the high-level properties of Chord.
We have first evaluated our implementation by performing 10,000 DHT `lookup'
requests generated from random nodes in the network for a random set of keys.
The lookups were generated after waiting for five minutes after the last node
joined in order to let the network stabilize.

\begin{figure}
\includegraphics{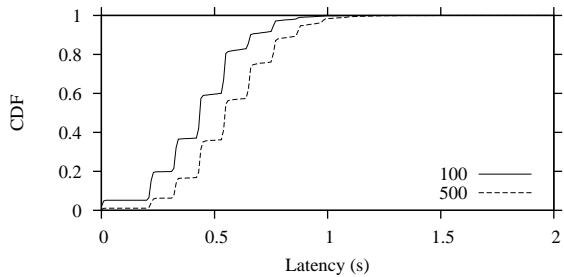}
\caption{Chord: Lookup latency distribution.}
\label{fig:chord_lookup_cdf}
\end{figure}

In Figure~\ref{fig:chord_lookup_cdf}, we present the cumulative distribution of
latency incurred to receive the response to the lookup requests with 100 and 500
nodes. The results are comparable or better than the published results for P2
Chord~\cite{LooSIGOPS05}.

\begin{figure}
\includegraphics{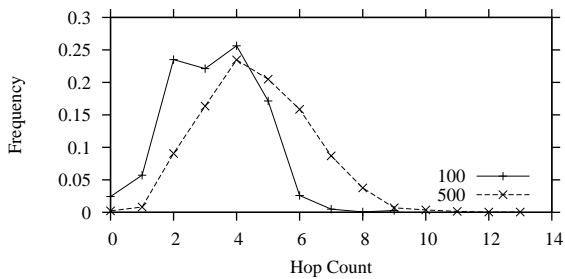}
\caption{Chord: Hop count distribution.}
\label{fig:chord_hop_count}
\end{figure}

In Figure~\ref{fig:chord_hop_count}, we present the frequency distribution of
the number of hops taken to complete the lookups.
As expected, the maximum number of hops taken is under
the theoretical upperbound of $\lceil\log{N}\rceil$.

\paragraph{Dynamic Membership} Our implementation of Chord in \DAHL also
handles churn.
In this experiment, we used 500 nodes, each one maintaining four successors and
performing finger fixing every 10 seconds and successor stabilization every 5
seconds.
This  configuration is similar to the setup of \Ptwo Chord.
We generated artificial churn in our experiment by killing and joining nodes at
random with different session times by following the methodology presented in
\cite{HandlingChurn}.

We obtained lookup consistency of 96\% for average session times of 60 minutes,
which is comparable with other implementations of Chord.

\paragraph{Summary} Our results show that our implementation of Chord in \DAHL
covers the  major algorithmic aspects of the protocol and that its run-time
performance is competitive with \Ptwo Chord.

\subsection{Zyzzyva}

\begin{figure}

\begingroup\small
\centering
\setlength{\fboxsep}{1.5pt}
\def\kw#1{\underline{\bfseries #1}}
\def\n{\string\+\ }
\def\lcolumn{0.55\textwidth}
\def\rcolumn{0.35\textwidth}
\def\lnhack{\gdef\lnhack{\\[0.5pt]}}
\newlength{\mycodeindent}\setlength{\mycodeindent}{2ex}
\newcounter{lineno}
\newcommand\Label[2][0]{%
  \lnhack\ttcode\refstepcounter{lineno}\label{#2}\thelineno:
  &\ttcode\hspace*{#1\mycodeindent}\ignorespaces}
\newcommand\State[1][0]{%
  \lnhack\ttcode\refstepcounter{lineno}\thelineno:
  &\ttcode\hspace*{#1\mycodeindent}\ignorespaces}
\def\Comment#1{&\multicolumn{1}{|p{\rcolumn}}{\itshape #1}}
\begin{tabular}{r@{}p{\lcolumn}p{\rcolumn}}
\State :- dynamic seqno/1, pending/3, cache/4. \Comment{Program state, and}
\State :- event request/1, process/2.          \Comment{event declarations.}
\State 
\Label{ln:request} request(Req) :-
                             \Comment{When I get a request \dots}
\State[1] \kw{this\_node}(ThisAddr),       \Comment{Find my own address,}
\State[1] primary(ThisAddr),          \Comment{if I'm the primary,}
\Label[1]{ln:sgn}\kw{signed\_by}(Src), \Comment{and got a signed request,}
\Label[1]{ln:n1}\n pending(\_, Src, Req), \Comment{which I haven't seen before,}
\State[1] count(pending(\_, \_, \_), N),{\rmfamily $^\dagger$} 
                                          \Comment{count the previous requests,}
\State[1] Id is N + 1,                    \Comment{to produce a new id, and}
\State[1]  assert(pending(Id, Src, Req)), 
                                 \Comment{add the new request as pending.}
\State[1]  batch\_size(Size),    \Comment{Recall the size of a batch,}
\State[1]   Id =:= Size,         \Comment{if there are enough requests,}
\State[1]   start\_new\_batch.
                               \Comment{start the protocol for this batch.}
\State     
\Label{ln:startbatch} start\_new\_batch :-      \Comment{When starting a new batch \dots}
\Label[1]{ln:fna}   findall(        
\State[2]       (Id, Src, Req),  \Comment{Collect all the pending requests,}
\Label[2]{ln:pop}  retract(pending(Id, Src, Req)), 
                                \Comment{removing them from the store,}
\State[2]       Batch           \Comment{and group them in a batch.}
\State[1]   ),
\State[1]   retract(seqno(Seq)),  \Comment{Get the next sequence number.}
\Label[1]{ln:sna}   \kw{sendall\_signed}(
\State[2]       Node,           \Comment{Ask all nodes,}
\State[2]       replica(Node),  \Comment{that happen to be replicas,}
\State[2]       process(Batch, Seq)
                             \Comment{to process this batch.}
\State[1]   ),
\State[1]   Next is Seq + 1, \Comment{Increment the sequence no.,}
\State[1]   assert(seqno(Next)). 
                           \Comment{and store the new value.}
\State     
\Label{ln:process}  process(Batch, Seq) :-
                        \Comment{When processing a batch \dots}
\State[1]   primary(Primary),         \Comment{Look up who is the primary,}
\State[1]   \kw{signed\_by}(Primary), \Comment{as this should be the one asking.}
\Label[1]{ln:all}   findall(\_, (
\State[2]     member((Id, Src, Req), Batch), \Comment{Unpack the batch,}
\State[2]     process\_req(Seq-Id, Src, Req)     
                                \Comment{and process each request.}
\State[1]   ), \_). 
\State     
\Label{ln:procreq} process\_req(Seq, Src, Req) :- 
                                \Comment{When processing a request \dots}
\Label[1]{ln:ite}   ( cache(Seq, Src, Req, Out) ->
                                \Comment{If I've seen this request before}
\Label[2]{ln:snd}     \kw{send\_signed}(Src, reply(Seq, Req, Out))
                                \Comment{reply with the cached response.}
\State[1]   ;              \Comment{Otherwise,}
\Label[2]{ln:n2}     \n cache(Seq, \_, \_, \_)),
                            \Comment{if it's a new sequence no.,}
\State[2]     compute\_output(Req, Out), \Comment{compute the output,}
\Label[2]{ln:add} assert(cache(Seq, Src, Req, Out)), \Comment{store it on the cache,}
\State[2]     \kw{send\_signed}(Src, reply(Seq, Req, Out)) 
                            \Comment{and send it back to the client.}
\State[1]   ). 
\end{tabular}
\endgroup

\caption{Initial phase of Zyzzyva with batching optimization.}

\bigskip
{\raggedright\noindent\footnotesize
  $^\dagger$\texttt{count/2} is a non-standard Prolog extension that counts the
  number of solutions of a given goal.\par}
  
\label{fig:zyzzyva}
\end{figure}

In this subsection, we evaluate the implementation of Zyzzyva in \DAHL.
For reference, and to give a flavor of the code written in \DAHL,
we include a fragment of the implementation of its first phase in
Figure~\ref{fig:zyzzyva}.
We present the peak throughput for the normal case, and the throughput after
killing a backup replica.
The goal of our experiments is 
to show that our implementation covers a significant part of Zyzzyva protocol and to show that its
performance is reasonable. We compare the performance of \DAHL Zyzzyva with the publicly available
C++ implementation of Zyzzyva (available from
\url{http://cs.utexas.edu/~kotla/RESEARCH/CODE/ZYZZYVA/}).

\paragraph{Setup} We use four physical machines as servers to tolerate one Byzantine faulty server and vary
the number of clients to measure the peak throughput. Both the server and client machines have identical
characteristics as previous experiment. The clients send requests with an empty payload, the execution cost of each operation
at the servers is zero, and we measure the peak throughput sustained by the servers. 

\paragraph{Implementation} We use OpenSSL's HMAC+MD5 cryptographic hash function in \DAHL to
perform critical digest and signing operations.  
Our implementation uses TCP as the transport protocol, we do
not yet use network broadcast feature, and do not implement batching.  Our implementation takes checkpoint at the rate of every 128 requests, which is 
standard in existing implementations. We do not implement state
transfer mechanism to bring the slow replicas up-to-date.

\paragraph{First case performance} In this experiment, we present the peak throughput of Zyzzyva without failures 
where requests are completed in single phase.  This 
result serves to measure the baseline functionality of Zyzzyva. The results are presented in 
Table~\ref{table:zyzzyva_performance_comparison}. We observe that the performance of \DAHL's Zyzzyva
is about~10 times slower than the C++ implementation.
However, as \citeN{clement09making} observe, the
penalty of using \DAHL over C++ will diminish as the application-level overhead starts to dominate. For example, with
an application that consumes approximately 100 $\mu$s per operation, Zyzzyva will deliver throughput of 9$\,$k req/s while
the implementation in \DAHL will deliver approximately 3$\,$k req/s, bringing down the penalty to 3X.

\paragraph{Second phase performance} In this experiment, we present the peak throughput of Zyzzyva 
when upto $\constF$ replicas are faulty and prevent requests from completing
in the single phase. This requires client to initiate the second phase, requiring more computation and
network resources at the replicas, resulting in lower performance compared to the previous result based
on single phase. Again, our results show that \DAHL implementation is slower compared to its 
counterpart in C++ owing to a slower runtime. 

\begin{table}
\caption{Zyzzyva: single phase and two phase performance for empty payload.}
\begin{tabular}{ccc}
\hline
\hline
& \DAHL Zyzzyva  & C++ Zyzzyva\\
\hline
Single phase & 4.5$\,$k req/s & 40$\,$k req/s \\
Second phase & 2.5$\,$k req/s & 20$\,$k req/s \\
\hline
\hline
\end{tabular}
\label{table:zyzzyva_performance_comparison}
\end{table}

\paragraph{Summary} The primary goal of our evaluation was to check if our implementation is comprehensive and faithful, and
to evaluate its performance. Our results show that the current implementation covers a significant portion of the protocol
features but the performance is lower compared to C++ implementation.  

\subsection{D'ARMC}
\begin{figure}
\includegraphics[scale=1.0]{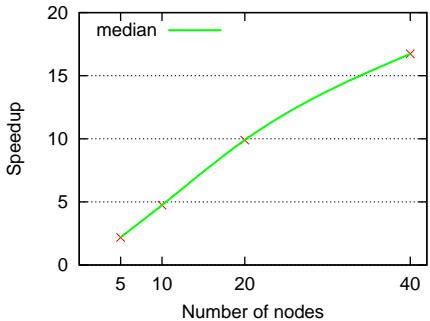}
\caption{Median speedup of D'ARMC with varying number of nodes.}
\label{fig:darmc-speedup}
\end{figure}

D'ARMC~\cite{darmc} is a distributed software model checker that was implemented
in \DAHL.
D'ARMC is a CPU-bound application, and therefore shows that \DAHL can be used
to implement more applications than mere network protocols.
The median speedup achieved by D'ARMC in a set of benchmarks is shown in
Figure~\ref{fig:darmc-speedup}.
The benchmarks consist in a set of automata-theoretic models from the
transportation domain and a standard hybrid-system example.

As can be seen in Figure~\ref{fig:darmc-speedup}, D'ARMC shows a linear speedup
with a varying number of machines, and the efficiency is about 50\%.
A more extensive evaluation can be found in~\cite{darmc}.

\subsection{Code Size} 
Our implementations of both Chord and Zyzzyva are comparable in size to the \Ptwo implementations in terms of lines
of code (LoC). For example, \DAHL Chord is implemented in 215 LoC while the \Ptwo Chord is implemented in 211 LoC. 
These sizes are an order of magnitude more succinct compared to a C/C++ implementation.


\section{Related work}
\label{sec:related}

In the previous section we have already compared DAHL with
two other related systems that help
the programmer to build distributed applications,
P2~\cite{LooSIGMOD06} and \Mace~\cite{KillianPLDI2007}.
Both languages have been successfully used for the implementation of
important networked systems and protocols, and serve as a research
platform for the development of specialized
variants --- see \url{http://declarativity.net} for further
pointers --- as well as verification
tools~\cite{MaceMCLive,CrystallBall,P2Sems,P2PADL,Cardan}.

Alternative approaches that attempt to extend Datalog for use in
a distributed environment, while trying to overcome the pitfalls 
of early Declarative Networking implementations, are
Meld \cite{Ashley07,Ashley09}, WIND \cite{Mao09} and Netlog \cite{Netlog}.
A common feature of these projects is that they all
argue that a `pure' Datalog based language is not appropriate
for the development of stateful applications. The authors of Meld
show that a limited declarative language can be used to program behavior
in ensembles; the authors of WIND propose the use of syntactic
`salt' to discourage, but still allow, the use of imperative
features; while the authors of Netlog augment Datalog rules with
annotations to explicitly control whether tuples are stored or sent
over the network.

In the broader picture of designing high-level languages for
concurrent and distributed programming, a prime example is
Erlang \cite{Erlang}. Erlang is based on the functional programming
paradigm and, similar to our approach, incorporates distribution
via explicit message passing between processes.
A related approach suggests using the Lua programming language
to implement distributed systems \cite{Leonini09}.

Some projects also aim to exploit the use of functional
programming languages at lower layers of the network protocol deign.
Foxnet, for example, implements the standard
TCP/IP networking protocol stack in ML \cite{Foxnet}; while
Melange provides a language to describe Internet packet protocols,
and generates fast code to parse/create these packets \cite{Melange}.
Similarly, the KL1 logic based language has been used to model
and exploit physical parallelism in the PIM operating system \cite{KL1}.

Previous work has also explored the use of Prolog to deal with
concurrency and parallelism, a comprehensive review is given
by \citeN{Gupta01}.
Most of this work, however, deals with the
problem of using Prolog to paralellize an otherwise sequential
task. Recent advances in this direction are discussed by
\cite{Casas08}.
Our work explores, instead, the use of Prolog as a general
purpose programming language to implement distributed applications.

\section{Conclusion}
\label{sec:conclusion}
\label{sec:discussion}

From our experience with applying Prolog for distributed
programming we draw the following conclusions.

In combination with event-driven control and networked communication
primitives, Prolog offers a programming language that is sufficiently
expressive and well-suited for the implementation of distributed
protocols.
In our experiments, we did not rely on any C/C++ extensions as there
was no need to compensate absence of certain programming constructs, as
it is common for the P2 system for declarative networking that is
Datalog-based.
Instead, we used the data type, control structures, and the database
facility provided by Prolog. 
By using Prolog as a basis we avoided any major
compiler/runtime/libraries implementation efforts, which often become
an obstacle when implementing a new programming language.
By not starting from scratch and relying on the existing Prolog
infrastructure, we obtain a fully-featured programming environment for distributed systems out of the box.


\bibliography{biblio}

\end{document}